\documentclass[final,5p,times,twocolumn]{elsarticle}

\usepackage{graphicx}

\usepackage{amssymb}

\journal{Nuclear Instruments and Methods A}

\begin{document}

\begin{frontmatter}



\title{Ageing studies of resistive Micromegas detectors for the HL-LHC}



\author[a]{J. Gal\'an}
\author[a]{D. Atti\'e}
\author[a]{E. Ferrer-Ribas}
\author[a]{A. Giganon}
\author[a]{I. Giomataris}
\author[a]{S.~Herlant}
\author[a]{F.~Jeanneau}
\author[a]{A.~Peyaud}
\author[a]{Ph. Schune}
\author[b]{T.~Alexopoulos}
\author[c]{M. Byszewski}
\author[b]{G.~Iakovidis} 
\author[d]{P. Iengo}
\author[b]{K. Ntekas}
\author[b]{S. Leontsinis}
\author[c]{R. de Oliveira} 
\author[b]{Y.~Tsipolitis}
\author[c]{J.~Wotschack}
\address[a]{IRFU, CEA-Saclay, 91191 Gif-sur-Yvette, France}
\address[b]{National Technical University of Athens, Zografou Campus, GR15773, Athens, Greece}
\address[c]{CERN, Geneva, Switzerland}
\address[d]{INFN, Napoli, Italy}

\begin{abstract}
Resistive-anode Micromegas detectors are in development since several years, in an effort to solve the problem of sparks when working in high flux and high radiations environment like in the HL-LHC (ten times the luminosity of the LHC). They have been chosen as one of the technologies that will be part of the ATLAS New Small Wheel project (forward muon system). An ageing study is mandatory to assess their capabilities to handle the HL-LHC environment on a long-term period. 
A prototype has been exposed to several types of irradiations (X-rays, cold neutrons, $^{60}$Co gammas) up to an equivalent HL-LHC time of more than five years without showing any degradation of the performances in terms of gain and energy resolution. Beam test studies took place in October 2012 to assess the tracking performances (efficiency, spatial resolution,...). Results of ageing studies and beam test performances are reported in this paper.
\end{abstract}


\end{frontmatter}


\section{Introduction}

High amplification gains are required in Micro-Pattern Gaseous Detectors (MPGD) in order to achieve a comfortable signal to noise ratio, therefore improving the performance of detectors in terms of spatial resolution and efficiency. The dense electron avalanches achieved at high gains with few primary electrons entail the risk of producing a discharge at the cathode of the detector when the critical electron density of $\sim 10^7 -  10^8$ electrons per avalanche is reached (related to the Raether limit~\cite{raether}). 

Discharges might affect the detector response in different ways;

\begin{itemize}
\item \emph{reducing its operating lifetime} due to intense currents produced in short periods of time, heating and melting the materials at the affected regions,
\item \emph{damaging the readout electronics} which have to support huge current loads, 
\item \emph{increasing the detector dead-time} resulting from a \emph{discharge of the cathode}\footnote{In this case, the amplification field is lost during a relatively long period of time due to the time required by the high voltage power supply to restore the charges, leading to an unavoidable detector dead-time}.
\end{itemize}

It was first observed in RPC-type detectors that the introduction of a high impedance resistive coating at the anode limits the detector current, constraining the spark process to the streamer phase during a time interval of at least some microseconds\footnote{The field is required to be lost for at least few microseconds allowing the high density electron and ion clouds to be evacuated before recovering field sparking conditions in the affected region.}. Thus reducing the total amount of charge released~\cite{fonte}. Furthermore, the limited discharge current affects the field locally, thus reducing the dead-time of the detector which remains operative in non-affected regions.

Micromegas detectors were introduced in 1996~\cite{mms} as a good candidate for high particle flux environments, and spark studies with detectors based on micromegas technology were also carried out~\cite{sparks2,sparks}. Recently, additional efforts have pushed the development of resistive strip micromegas detectors in order to further increase its robustness in high particle flux environments by limiting spark discharges in the same way as it was done for RPCs. In particular, the MAMMA collaboration~\cite{mamma} is developing large area micromegas detectors and introduced the resistive coating technique~\cite{largeMM} for the upgrade of the HL-LHC\footnote{High Luminosity Large Hadron Collider (luminosity will be increased by at least a factor 5 reaching up to $L = 5\times10^{34}$\,cm$^{-2}$s$^{-1}$)}.

The existing Micromegas technology~\cite{bulk} allowed the MAMMA collaboration to investigate detector prototypes with different resistive coating topologies~\cite{resist}. This new type of detectors will be installed in ATLAS at the New Small Wheel project. Thus, it should be proved to be long term radiation resistant. The introduction of a new technology made of new materials adds some uncertainties of operation during long periods of time in intense particle flux environments. The results that we report here represent the first ageing tests with this type of detectors using different types of highly ionizing radiations.



\section{Ageing of resistive micromegas detectors}\label{sc:ageing}

An ageing study of resistive-anode detectors is required to operate these type of detectors at the HL-LHC. For this study, two new identical micromegas prototypes (named R17) were used, lent by the MAMMA collaboration and built at the CERN workshop.

These detectors are based on a resistive strips technology \cite{resist} with a 2-dimensional readout. The X and Y readout strips are in copper. The top Y strips have been covered by a 60\,$\mu$m thick insulating coverlay. The 35\,$\mu$m thick resistive strips are placed on top of this layer parallel to the X strips.

Both resistive and copper strips have a pitch of 250\,$\mu$m and a width of 150\,$\mu$m. The resistivity along the strips and boundary resistance value was measured during the fabrication process. The first detector, R17a, showed a linear resistivity of 45-50 M$\Omega$\,cm$^{-1}$, and a boundary resistance of 80-140 M$\Omega$. The resistivity obtained for the second detector, R17b, was comparable with a linear resistivity of 35-40 M$\Omega$\,cm$^{-1}$ and a boundary resistance of 60-100 M$\Omega$.

\begin{figure}[htp]
\begin{center}
\includegraphics[width=0.8\columnwidth,keepaspectratio]{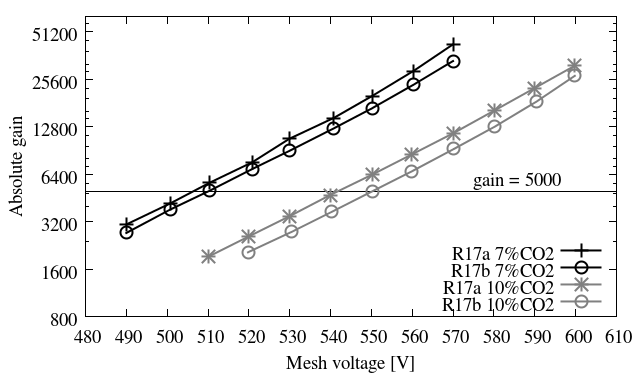}
\caption{ Gain curves of each detector for two different gas mixtures, argon as main component.}
\label{resist_princ}
\end{center}
\end{figure}

A first characterization in Ar+CO$_2$ mixtures\footnote{The mixture used during ageing periods described in the text was always Ar + 10\% CO$_2$, and Ar + 7\% CO$_2$ for the test beam performance.} showed that the gain behaved as expected, and the results obtained with both detectors, R17a and R17b, were comparable. Figure~\ref{resist_princ} shows the gain curves in these mixtures.

Ageing tests took place at different CEA-Saclay facilities, one prototype (R17b) was kept unexposed as a reference and the other (R17a) was exposed to different types of radiations including X-rays, cold neutrons, high energetic gammas, and alphas. These tests will be described in the following sections.

\subsection{X-ray exposure}

The prototype under test was placed inside a cage with a high intensity X-ray generator, which emits in all the X-range up to few keV, with a copper fluorescence at 8\,keV. A mask was prepared by using an aluminum plate with a 4 cm$^2$ hole aperture. The plate was placed on top, and it was always fixed in the same position.


The radiation exposure tests aim to accumulate an amount of charge comparable to the values that will be integrated during the lifetime of the HL-LHC. The estimation of the total charge produced at the HL-LHC in the muon chambers of ATLAS is based on the energy deposit $E_{MIP}$ of a Minimum Ionizing Particle (MIP) in a 0.5\,cm micromegas conversion gap. In our gas mixture $E_{MIP} = 1.25 $ keV. Considering the detector gain $G$, the charge produced by each incident particle at the HL-LHC

$$
Q_{MIP}=\frac{E_{MIP}}{W_i} q_e G
$$

\noindent where $W_i = 26.7$ eV is the mean ionization energy of the gas. Taking into account a nominal gain of G=5,000, the charge produced per MIP yields $Q_{MIP} = 37.4$ fC.

Assuming the expected rate at the HL-LHC future muon chambers will be 10 kHz/cm$^2$, the total charge generated in 5 years of operation (200 days/year) will be 32.3 mC/cm$^2$. The flux produced at the X-ray generator will accumulate an amount of charge well above this value for an exposure of a few days.

\begin{figure}[ht!]
\begin{center}
\includegraphics[width=\columnwidth,keepaspectratio]{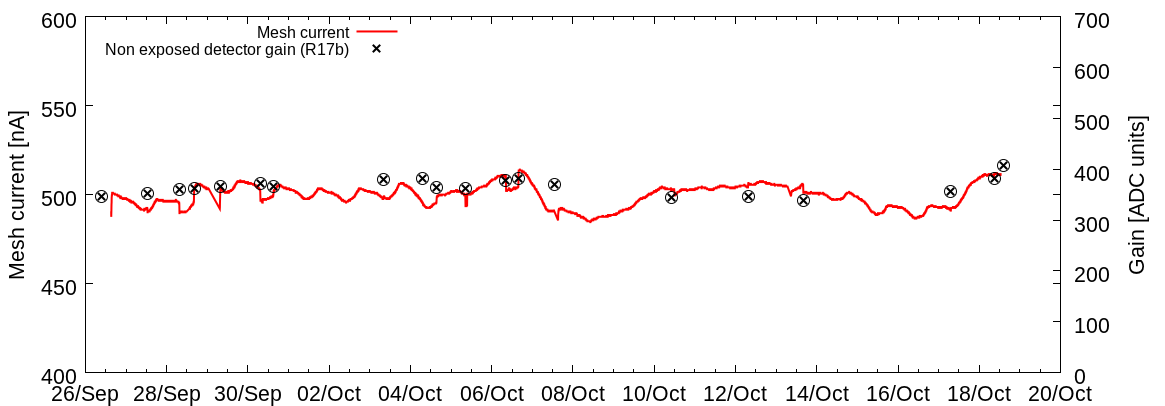}
\caption{Mesh current evolution (red curve) for a period of 21 days and an integrated charge of 918 mC. The gain control measurements with the non-exposed detector (which was connected in the same gas line in parallel) are also plotted (black circles).}
\label{xray_current}
\end{center}
\end{figure}

The detector has been exposed for more than 20 days with a gain of 5,000 and a gas flow of one renewal per hour. The current remained stable (see figure~\ref{xray_current}), proving the detector gain was not affected on the whole irradiation period, which corresponds to 21.3 days of exposure and an integrated charge of 918 mC, that is 5 years of HL-LHC with a security factor above 7.

\begin{figure}[htbp]
\begin{center}
\includegraphics[width=0.7\columnwidth,keepaspectratio]{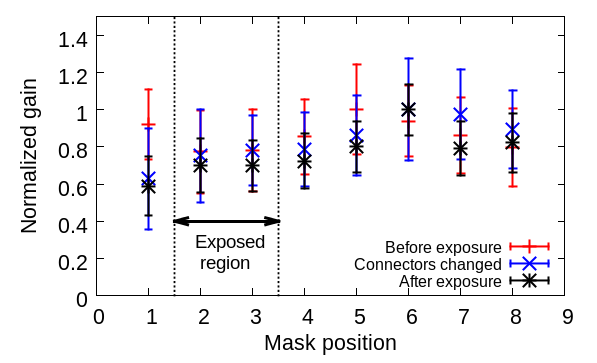}
\caption{Gain measurements as a function of position before, during (connectors changed, details can be found at reference~\cite{javier}), and after the irradiation period for the exposed detector R17a (Left) and non-exposed detector R17b (Right). The exposed region is indicated for the irradiated detector.}
\label{gain_profile}
\end{center}
\end{figure}

The gain of both detectors was measured using an $^{55}$Fe source at different positions, before and after exposure. We used a dedicated 9-hole mask which covers the full active area of the detector in one of its axis. These measurements took place before the ageing period (26-Sep-2011), when the grounding connectors were removed (8-Oct-2011) and after the exposure (19-Oct-2011) (further details can be found at references \cite{javier,fabien}). The relative gain at each position for these three sets of measurements is plotted in figure~\ref{gain_profile} and shows that the gain profile is compatible with previous measurements. Moreover, the exposed detector region does not show any significant difference in relative gain compared to the other, non-exposed regions. The X-rays irradiation had therefore no effect on the detector response.

\subsection{Neutron exposure}

Neutron irradiation took place at the Orph\'ee reactor in CEA-Saclay. The reactor, which operates only for research purposes, is connected to different lines which guide different fluxes of cold neutrons produced in the reactor. The line where the detector was installed provides a neutron beam of about 8$\times$10$^8$\,cm$^{-2}$s$^{-1}$, with energies in the range of 5-10\,meV within an area of a few cm$^2$.


%

\begin{figure}[htbp]
\begin{center}
\includegraphics[width=\columnwidth,keepaspectratio]{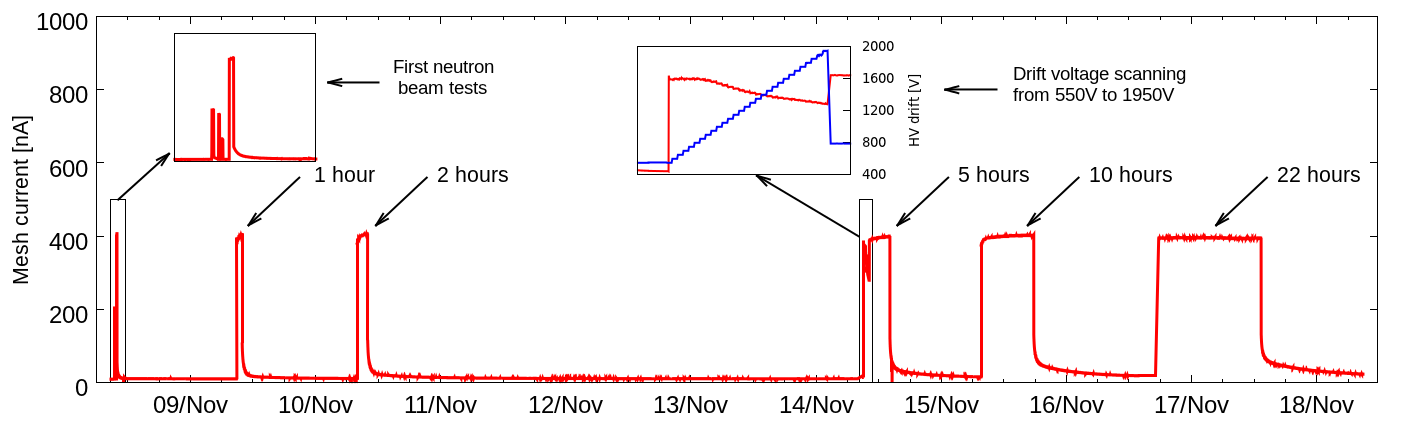}
\caption{Mesh current during the different neutron irradiation periods. The drift voltage was modified during the 5 hours period. Thus, proving the mesh current dependency with the mesh electron transmission when the neutron beam was active (related to the micromegas field ratios~\cite{paco}).}
\label{neutron_current}
\end{center}
\end{figure}

Different irradiation periods were scheduled with an increased time exposure. The mesh current remained stable and at the same level in each of these periods (see figure~\ref{neutron_current}).

The expected neutron flux at the CSC (Cathode Strip Chamber) in ATLAS is about 3$\times$10$^4$ cm$^{-2}$s$^{-1}$. The total exposure time of the prototype R17a was more than 40 hours, accumulating a total amount of neutron flux which is equivalent to 5 years of operation of the HL-LHC with a safety factor well above~10. 

\begin{figure}[htbp]
\begin{center}
\includegraphics[width=0.7\columnwidth,keepaspectratio]{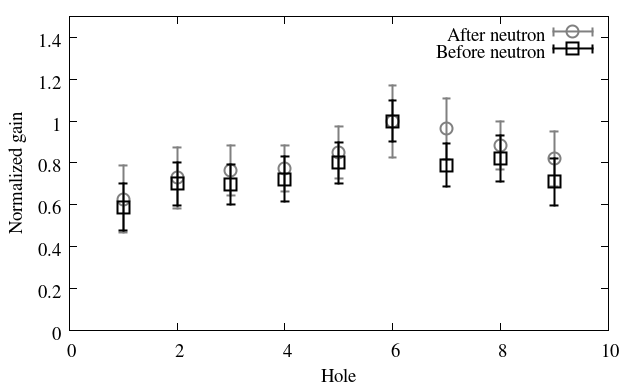}
\caption{Measured gain profile before and after neutron irradiation (error-bars are proportional to the energy resolution).}
\label{neutron_profile}
\end{center}
\end{figure}

Before and after the neutron tests the gain was measured using the same 9-hole mask used for the tests described in the previous section. The gain profile is compatible and the performance of the detector shows no degradation with respect to the measurements before neutron irradiation (see figure~\ref{neutron_profile}).

\subsection{Gamma exposure}

After proving that the detector R17a was operating properly it was installed at the COCASE~\cite{COCASE} gamma facility, which provides a high activity $^{60}$Co source, of about 500 mGy/h, emitting gammas at 1.17 MeV and 1.33 MeV. 


The highest activity for gammas at the muon spectrometer is recorded in the forward CSC region~\cite{bkg}, with a flux which is below 1.8$\times$10$^4$\,cm$^{-2}$s$^{-1}$. For 5 years of running HL-LHC, we consider a factor 5 in luminosity increase, with a safety factor 3 the integrated gamma flux results to be 2.3$\times$10$^{13}$\,cm$^{-2}$.

The detector was placed at 50\,cm from the source receiving an equivalent flux of 1.7$\times$10$^7$\,cm$^{-2}$s$^{-1}$ which should be uniformly distributed in the active area of the detector. Thus, time required to reach the expected gamma flux integrated for 5 years of HL-LHC would be of 16 days in COCASE.

\begin{figure}[htbp]
\begin{center}
\includegraphics[width=\columnwidth,keepaspectratio]{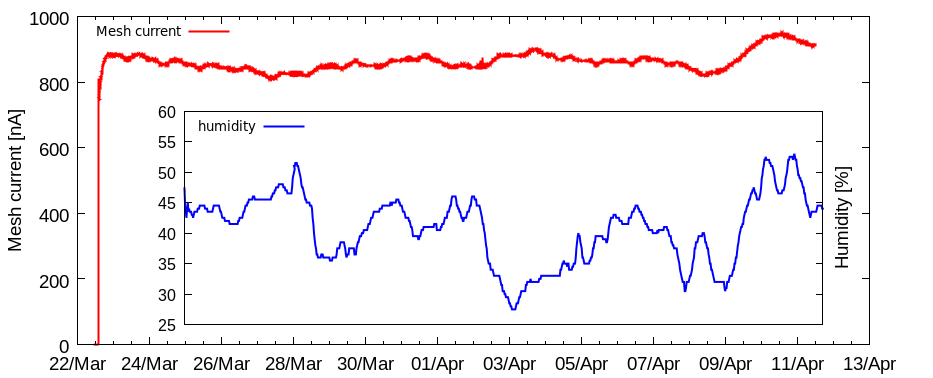}
\caption{Mesh current evolution during the gamma irradiation period with a zoomed plot inside showing humidity measurements taken at the COCASE facility during these tests.}
\label{gamma_current}
\end{center}
\end{figure}

The detector was irradiated during 480 hours between March,22$^{nd}$ and April,11$^{th}$ 2011. The integrated charge during this period was 1.48\,C at a mean mesh current of 858 nA. Figure \ref{gamma_current} shows the evolution of the mesh current which fluctuates around the mean value within a 5\%, variation which can be perfectly attributed to environmental effects.

\begin{figure}[htbp]
\begin{center}
\includegraphics[width=0.7\columnwidth,keepaspectratio]{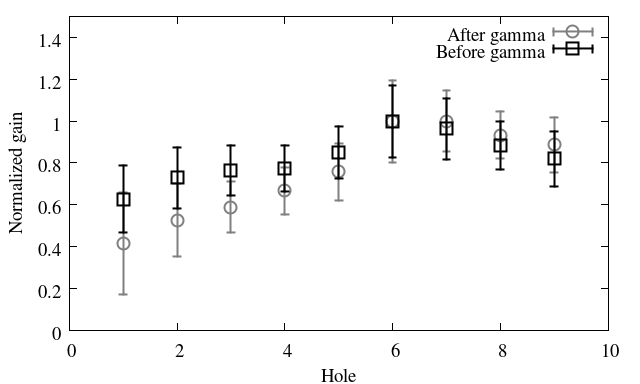}
\caption{ The 9-hole mask normalized gain profile before and after gamma irradiation (error-bars are proportional to the main $^{55}$Fe peak energy resolution).
}
\label{gamma_profile}
\end{center}
\end{figure}

The gain profile before and after exposure shows a reasonable reproducibility (see figure~\ref{gamma_profile}).

\section{Beam test performance after irradiation}

After the completion of the above described exposures, the R17a and R17b prototypes were installed in the H6 CERN-SPS beam line. The beam consists of positively charged pions at 120\,GeV/c in spills of about 10\,s every 48\,s. The beam intensity was typically 5$\cdot$10$^4$/spill over an area of 20\,mm$\times$10\,mm. The DAQ recorded around 800 events per spill.

\begin{figure}[htbp]
\begin{center}
\includegraphics[width=0.8\columnwidth,keepaspectratio]{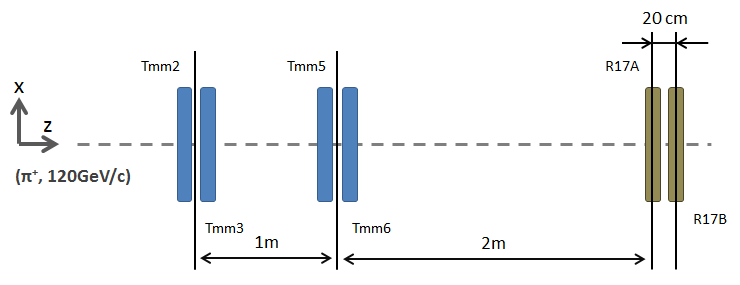}
\caption{ Test bench setup used at the CERN pion beam facility. Beam is coming from the left and goes through reference telescope detectors Tmm$_n$. The line is shared with other detectors which were under test at the same time and have been excluded in this drawing for simplicity. Our resistive prototypes were placed behind the telescope. }
\label{beam_setup}
\end{center}
\end{figure}

The intention of this last test was to determine any hint of ageing of the R17a by comparing the spatial resolution (SR) and efficiency obtained with respect to the reference detectors. The R17a and R17b were installed in the existing micromegas telescope line, as shown in figure~\ref{beam_setup}. The micromegas telescope (Tmm2, Tmm3, Tmm5, Tmm6) is used as reference for the track reconstruction of the beam. The detectors under test were placed 2\,m downstream of Tmm6.

The data were taken within two weeks at the end of October 2012. Three different regions were exposed to the beam in order to compare zones which received different types of radiations. We took data at each of these regions for several values of the amplification field by varying the mesh voltage, observing no significant difference between zones.

\begin{figure}[htbp]
\begin{center}
\includegraphics[width=0.49\columnwidth,keepaspectratio]{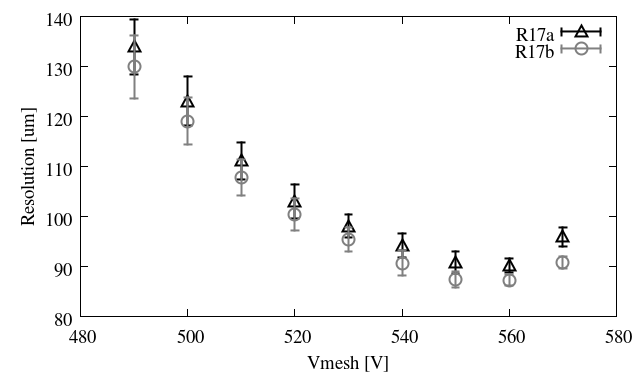}
\includegraphics[width=0.49\columnwidth,keepaspectratio]{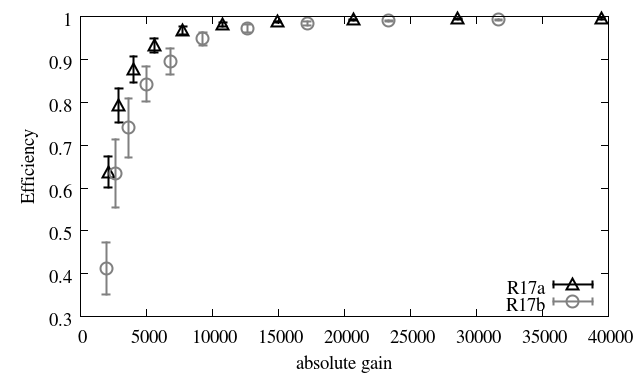}
\caption{ (Left) Comparison of the SR of R17a and R17b as a function of the mesh voltage. (Right) Comparison of R17a and R17b average efficiencies as a function of the absolute gain.}
\label{efficiency}
\end{center}
\end{figure}

Figure~\ref{efficiency} (Left) shows the averaged SR as a function of the mesh voltage pointing to an optimum SR at about 550\,V, which corresponds to gains slightly above 10,000. Figure~\ref{efficiency} (Right) shows the detection efficiency of both detectors as a function of the absolute gain. The efficiencies were determined as the probability to detect one charge cluster when a track is found in each of the \emph{four} telescope detectors. Both detectors reach efficiencies of about 99.5\% for the highest values of the gain, proving that there is no visible degradation effect in these measurements (further details can be found at~\cite{ageingPaper}).

\section{Conclusions}

A 2D-readout resistive strip detector was tested to different intense irradiation sources: X-rays, high energy gammas, neutrons and alphas~\cite{ageingPaper}. For each radiation type, the total charge or interactions produced during the ageing process was well above the expected levels at the HL-LHC for 5 years of operation with an additional safety factor. In addition we investigate the performance of these detectors to the CERN pion beam test facility. The results obtained are comparable to those obtained with a non-irradiated detector of same construction. We did not observe any degradation. The intense irradiation of R17a detector did not harm seriously the performance of the detector in terms of efficiency and SR, similar to the reference telescope. We present it as a definitive proof that irradiations of different nature, which will be present at the HL-LHC, do not affect this new technology. The values obtained for the spatial resolution and detection efficiency are reasonably good considering that the setup was not optimized to minimize the error on the track definition.

We consider this as an important step towards the consolidation of this technology for high-rate environments in long periods of time.








\begin{thebibliography}{00}

\bibitem{raether}
H. Raether, \emph{Electron avalanches and breakdowns in gases}, Washington: Butterworths, 1964. 

\bibitem{fonte}
P. Fonte {\em et al},
\emph{ A spark-protected high-rate detector},
{\emph{Nucl. Instr. and Meth. A}, {\bf 431} (1999) p. 154}.

\bibitem{mms}
Y. Giomataris {\em et al.},
\emph{MICROMEGAS: A High granularity position sensitive gaseous detector for high particle flux environments},
{\emph{Nucl. Instr. and Meth. A}, {\bf 376} (1996) p. 29}.

\bibitem{sparks2}
D. Thers {\em et al.},
\emph{Micromegas as a large microstrip detector for the COMPASS experiment},
{\emph{Nucl.\,\,Instr.\,\,and\,\,Meth.\,\,A}, {\bf 469} (2001) p. 133}.


\bibitem{sparks}
A. Bay {\em et al.},
\emph{Study of sparking in Micromegas chambers},
{\emph{Nucl.\,\,Instr.\,\,and\,\,Meth.\,\,A}, {\bf 488} (2002) p. 162}.

\bibitem{mamma}
{\emph{Muon Atlas MicroMegas Activity} website}

\bibitem{largeMM}
T. Alexopoulos {\em et al.},
\emph{Development of large size Micromegas detector for the upgrade of the ATLAS Muon system},
{\emph{Nucl. Instr. and Meth. A}, {\bf 617} (2010) p. 161}.

\bibitem{bulk}
I. Giomataris {\em et al.}, \emph{Micromegas in a bulk}, 
{\emph{Nucl. Instr. and Meth. A}, {\bf vol. 560} (2006), no. 2, p. 405}. 

\bibitem{resist}
T. Alexopoulous {\em et al.}, \emph{A spark-resistant bulk-micromegas chamber for high-rate applications},
{ \emph{Nucl. Instr. and Meth. A}, {\bf vol. 640} (2011), p. 110}. 

\bibitem{javier}
J. Gal\'an {\em et al.}, \emph{Aging studies of Micromegas prototypes for the HL-LHC}, \emph{2012 JINST 7 C01041}.

\bibitem{fabien}
F. Jeanneau {\em et al.}, \emph{Performances and Ageing Study of Resistive-Anodes Micromegas Detectors for HL-LHC Environment}, 
{\emph{Nuclear Science, IEEE Transactions on} , vol.59, no.4, pp.1711-1716, Aug. 2012}.

\bibitem{paco}
F.J. Iguaz {\em et al.}, \emph{Characterization of microbulk detectors in argon- and neon-based mixtures}, \emph{2012 JINST 7 P04007}.

\bibitem{COCASE}
R. Chipaux and O. Toson, \emph{Resistance Of Lead Tungstate And Cerium Fluoride To Low Rate Gamma Irradiation Or Fast Neutrons Exposure},
{CMS Technical Note CMS TN, 1995, p. 95-126.}

\bibitem{bkg}
S.~Baranov {\em et al.}, \emph{Estimation of Radiation Background, Impact on Detectors, Activation and Shielding Optimization in ATLAS},{\emph{ATL-GEN-2005-001.}}



%
\bibitem{ageingPaper}
J. Galan {\em et al.}, \emph{An ageing study of resistive micromegas for the HL-LHC environment}, submitted to JINST [arXiv:1301.7648].


\end{thebibliography}
\end{document}